\documentclass[aps,twocolumn,reprint,english, superscriptaddress,eqsecnum,showpacs,longbibliography]{revtex4-1}
\usepackage{amssymb}
\usepackage{amsmath}
\usepackage{graphicx}
\usepackage{xcolor}
\usepackage{epsfig}
\definecolor{darkblue}{rgb}{0.1,0.2,0.6}
\definecolor{darkred}{rgb}{0.8,0.1,0.2}
\usepackage[colorlinks,
            citecolor=darkblue,
            linkcolor=darkred,
            urlcolor=darkblue]{hyperref}
\usepackage[all]{hypcap}
\usepackage{braket}
\usepackage[normalem]{ulem}

\newcommand{\revise}[1]{{\color{black} #1}}
\newcommand{\secondrevise}[1]{{\color{black} #1}}
\newcommand{\thirdrevise}[1]{{\color{black} #1}}
\newcommand{\fourthrevise}[1]{{\color{black} #1}}

\makeatletter
\adddialect\l@English\l@english
\makeatother

\begin{document}

\title{Universal short time quantum critical dynamics of finite size systems }

\author{Yu-Rong Shu}
\affiliation{State Key Laboratory of Optoelectronic Materials and Technologies, School of Physics, Sun Yat-Sen University, Guangzhou 510275, China}
\author{Shuai Yin}
\affiliation{State Key Laboratory of Optoelectronic Materials and Technologies, School of Physics, Sun Yat-Sen University, Guangzhou 510275, China}
\affiliation{Institute for Advanced Study, Tsing Hua University, Beijing 100084, China}
\author{Dao-Xin Yao}
\email{yaodaox@mail.sysu.edu.cn}
\affiliation{State Key Laboratory of Optoelectronic Materials and Technologies, School of Physics, Sun Yat-Sen University, Guangzhou 510275, China}
\date{\today}

\begin{abstract}
We investigate the short time quantum critical dynamics in the imaginary time relaxation processes of finite size systems. Universal scaling behaviors exist in the imaginary time evolution and in particular, the system undergoes a critical initial slip stage characterized by an exponent $\theta$, in which an initial power-law increase emerges in the imaginary time correlation function when the initial state has zero order parameter and vanishing correlation length. Under different initial conditions, the quantum critical point and critical exponents can be determined from the universal scaling behaviors. We apply the method to the one- and two-dimensional transverse field Ising models using quantum Monte Carlo simulations. In the one-dimensional case, we locate the quantum critical point at $(h/J)_{c}=1.00003(8)$ \thirdrevise{in the thermodynamic limit}, and estimate the critical initial slip exponent $\theta=0.3734(2)$, static exponent $\beta/\nu=0.1251(2)$ \thirdrevise{by analyzing data on chains of length $L=32\sim 256$ and $L=48\sim 256$, respectively}. For the two-dimensional square-lattice system, the critical coupling ratio is given by $3.04451(7)$ \thirdrevise{in the thermodynamic limit} while the critical exponents are \thirdrevise{$\theta=0.209(4)$ and $\beta/\nu=0.518(1)$ estimated by data on systems of size $L=24\sim 64$ and $L=32\sim 64$, correspondingly.} Remarkably, the critical initial slip exponents obtained in both models are notably distinct from their classical counterparts, owing to the essential differences between classical and quantum dynamics. The short time critical dynamics and the imaginary time relaxation QMC approach can be readily adapted to various models.
\end{abstract}
\maketitle

\section{Introduction}
\label{sec:introduction}
The understanding of non-equilibrium dynamics in interacting many-body quantum systems is a crucial issue in modern physics with increasing focus stimulated by the experimental developments in the field of cold atoms \cite{bloch08}.
Different approaches can be used to take a system out of equilibrium, such as applying a driving field or pumping energy into the system, among which one simple but interesting candidate is to impose a sudden quench on an equilibrium system \cite{dziarmaga10,polkovnikov11}.
After the sudden quench, the system may go through different types of non-equilibrium processes, for instance, it can gradually relax back to equilibrium or enter a quasi-steady prethermal state and end up in a thermalized state \cite{berges04}, depending on the setup of the system and the dynamics that governs the evolution.
Experimental interests in the quantum quench dynamics are heating up \cite{sadler06,ritter07,trotzky12,navon15,nicklas15} and many theoretical efforts have also been dedicated, some of which focus on the postquench long time universal scaling behaviors \cite{lamacraft07,rossini09,venuti10,cdgrandi10,deng11,foini12,iyer12,karrasch12,karl17} while others pay their attention to the transient dynamics in the short time regime after the quench \cite{torre13,yin14,gagel14,gagel15,calabrese12_1,calabrese12_2,maraga15,chiocchetta15,chiocchetta16_quantum,chiocchetta17,wliu16}.

In the late 1980's, Janssen {\it et al} \cite{janssen89} and Huse \cite{huse89} discovered respectively the universal short time critical dynamics (STCD) in classical phase transitions.
Universal scaling behaviors are found in the relaxation process after performing a sudden quench to the critical point from high temperature with a small order parameter $M_{0}$ and vanishing correlation length \cite{janssen89,zbli95,zbli96}.
The emergence of the initial power-law increase in the order parameter, characterized as the critical initial slip with an independent exponent $\theta$, is an important feature of the STCD.
In the short time regime, after a transient microscopic period, the order parameter increases as $M(t)\propto M_{0}t^{\theta}$.
The characteristic time scale of this short time regime depends on $M_{0}$ and scales as $t_{\text{cr}}\sim M_{0}^{-z/x_{0}}$, in which $x_{0}=\theta z+\beta/\nu$ is the scaling dimension of $M_{0}$ \cite{janssen89,diehl93}.
After $t_{\text{cr}}$, the initial condition becomes irrelevant and the behavior of $M(t)$ crosses over to the well-known long time power-law decay $M(t)\sim t^{-\beta/\nu z}$ \cite{janssen89}.
In the past few decades, the STCD has become a powerful tool in studying critical properties \cite{zbli94,zbli95,zbli96,bzheng98} and successfully applied to various models \cite{hpying01,albano11,tome98,bzheng96,bzheng03,silva02,arashiro03,baumann07,silva13,silva14,schulke95,luo97,grassberger95}.

The classical STCD has been studied extensively (see Ref.~\cite{albano11} for a review) while the same issue in quantum systems is attracting increasing attention in recent years.
Experimentally, temporal scaling crossover is observed in the time evolution of an isolated one-dimensional (1d) two-component Bose gas, when quenched into the vicinity of the critical point \cite{nicklas15}. The critical initial slip and crossover to long time non-critical regime are found in the correlation length of the order parameter \cite{nicklas15}.
Theoretical interests focus on several aspects, including open and isolated quantum systems
as well as real and imaginary time evolutions \cite{calabrese12_1,calabrese12_2,gagel14,gagel15,maraga15,chiocchetta15,chiocchetta16_quantum,chiocchetta17,wliu16,yin14,zhang14,weinberg17}.
The works in Refs.~\cite{gagel14,gagel15} present investigations on an $N$-component $\varphi^{4}$ model coupled to an external bath, where the critical initial slip and long time power-law decay are found in the real time relaxation of quantum open systems.
A renormalization group analysis by Chiocchetta {\it et al} \cite{chiocchetta17} shows the existence of the critical initial slip and dynamical crossovers between different scaling regimes in prethermal states.
Depending on dimension and the energy scale injected by the quench, crossover can exist between two critical initial slip stages with different exponents $\theta$'s governed by a quantum and a classical prethermal fixed point correspondingly \cite{chiocchetta17}.
Due to the unitary nature of real time evolution, only the prethermal stage has similar evolution behaviors as the classical case \cite{maraga15,chiocchetta15,chiocchetta16_quantum,chiocchetta17}.
Differently, in imaginary time, high energy excited states decay so fast that the evolution is governed by low energy levels close to the ground state and exhibits universal power-laws \cite{yin14,zhang14}.
A scaling theory for the quantum STCD in imaginary time was proposed in Ref.~\cite{yin14}, and realized in 1d Ising systems.
The critical initial slip also appears in the imaginary time evolutions of quantum systems, i.e., the time dependence of the order parameter in Ref.~\cite{janssen89} also holds in imaginary time $M(\tau)\propto M_{0}\tau^{\theta}$ for a small finite $M_{0}$ \cite{yin14}.
The critical initial slip terminates at time scale $\tau_{\text{cr}}\sim M_{0}^{-z/x_{0}}$, which is called the initial-value time \cite{diehl93}, and crosses over to the universal long time power-law decay $M(\tau)\sim \tau^{-\beta/\nu z}$. The value of the critical initial slip exponent is found to be $\theta=0.373$ \cite{yin14} in the 1d transverse field Ising model~(TFIM), in sharp difference from its classical counterpart, the two-dimensional (2d) classical Ising model $\theta=0.191(1)$ \cite{okano97,albano11}.

Even with the progress made by density-matrix renormalization group and related matrix-product state methods \cite{schollwock05,schollwock11,daley04}, real time dynamics is difficult to deal with computationally for most of quantum systems, especially for higher dimensions. As demonstrated in Refs.~\cite{cdgrandi11,cdgrandi13,cwliu13,witczak14,karrasch12}, real time and imaginary time dynamics bear so many similarities that some scaling properties of real time dynamics can be predicted by the imaginary time dynamics. Recent developments in quantum Monte Carlo~(QMC) \cite{cwliu13,cdgrandi11,cdgrandi13} light up the possibility to simulate the imaginary time relaxation process in a wide range of ``sign-problem'' free quantum models, offering a bright way to study the quantum STCD.

In this work, we study the quantum STCD of finite size systems in imaginary time relaxation and apply it to both the 1d and 2d TFIMs using a ground state projection QMC approach.
The quantum critical point and critical exponents ($\theta$ and $\beta/\nu$) are extracted from the universal scaling behaviors.
The key result of our work is the capture of the critical initial slip of the 2d square-lattice TFIM. To our knowledge, the critical initial slip exponent $\theta$ of the 2d TFIM has not been computed yet. In the meantime, we obtain the $\theta$ of the 1d TFIM, soundly confirm the result found in Ref.~\cite{yin14}.
In both cases, the critical initial slip exponents are distinct from their classical counterparts.
Our results show the capability of the quantum STCD method in determining critical properties and shed light on future applications to other models.

The rest of the paper is organized in the following way. In Sec.~\ref{sec:img} we first briefly review the imaginary time dynamics. In Sec.~\ref{sec:ffs} we generalize the scaling theory for the quantum STCD to finite size systems in imaginary time. The key idea of the projector QMC method is outlined in Sec.~\ref{sec:method} while Sec.~\ref{sec:model} introduces the properties of the models studied. Numerical results of the 1d and 2d TFIMs are presented in Sec.~\ref{sec:results} and a summary is given in Sec.~\ref{sec:summary}.

\section{Imaginary time dynamics}
\label{sec:img}
In this section, we discuss the imaginary time evolution of a quantum state $\ket{\psi(\tau)}$.
By performing the standard Wick rotation of $t\rightarrow -i\tau$, the Schr\"{o}dinger equation describes the imaginary time evolution of the wavefunction as \cite{justin96,altland06}
\begin{equation}
  \label{eq:Seq}
  \frac{\partial}{\partial\tau}\ket{\psi(\tau)}=-H\ket{\psi(\tau)},
\end{equation}
with the Planck constant being $1$ and normalization condition $\braket{\psi(\tau)|\psi(\tau)}=1$.
A formal solution of Eq.~(\ref{eq:Seq}) is given by
\begin{equation}
  \label{eq:formal}
\ket{\psi(\tau)}=Z_{0}^{-1}e^{-\tau H}\ket{\psi(\tau_{0})},
\end{equation}
in which $e^{-\tau H}$ is the imaginary time evolution operator and the normalization factor $Z_{0}$ is
\begin{equation}
  \label{eq:norm}
  Z_{0}=||e^{-\tau H}\ket{\psi(\tau_{0})}||,
\end{equation}
where $||...||$ denotes a modulo operation.

In the energy representation, Eq.~(\ref{eq:formal}) reads
\begin{eqnarray}
  \label{eq:psi_expand}
  \ket{\psi(\tau)} &=& Z_{0}^{-1}\sum_{i}{c_{i}e^{-E_{i}\tau}\ket{i}} \nonumber \\
                   &=& Z_{0}^{-1}e^{-E_{0}\tau}\sum_{i}{c_{i}e^{-(E_{i}-E_{0})\tau}\ket{i}} \nonumber \\
                   &\sim& c_{0}\ket{0}+c_{1}e^{-\Delta \tau}\ket{1},
\end{eqnarray}
where the coefficients are given by the overlap of the initial state and the $i$th eigenstate of the Hamiltonian $c_{i}=\braket{i|\psi(\tau_{0})}$ and $\Delta=E_{1}-E_{0}$ is the gap between the first excited state and the ground state. It is indicated in Eq.~(\ref{eq:psi_expand}) that the high energy levels decay too fast to be non-negligible and $\ket{\psi(\tau)}$ is governed by the low energy levels during the imaginary time evolution \cite{yin14}.

As pointed out in Ref.~\cite{zhang14}, a dissipative equation of the probability to find a given eigenstate can be derived from the Schr\"{o}dinger equation and it is argued that this dissipative equation in fact exhibits similar evolution properties with the classical master equation, especially the critical initial slip in the short time stage, though, the dynamics are essentially different \cite{zhang14}.

\section{Quantum short time critical dynamics in finite size systems}
\label{sec:ffs}
In this section, we discuss the quantum STCD in finite size systems during the imaginary time evolution under different quench protocols.

When a system is initially off-critical with vanishing correlation length, a sudden quench to the critical point triggers the relaxation \cite{janssen89,yin14}. In imaginary time, universal scaling behaviors are found after a short period of non-universal microscopic time scale $\tau_{\text{mic}}$ \cite{yin14}. Taking the system size into account, the scaling form of an observable $P$ is given by \cite{yin14}

\begin{equation}
P(\tau,g,M_{0},L)=b^{\phi}P(\tau^{\prime},g^{\prime},M_{0}^{\prime},L^{\prime}),
\label{eq:pscaling}
\end{equation}
where $\tau^{\prime}=b^{-z}\tau$, $g^{\prime}=b^{1/\nu}g$, $M_{0}^{\prime}=b^{x_{0}}M_{0}$, and $L^{\prime}=b^{-1}L$ for an arbitrary scaling factor $b$.
The arguments $\tau$, $g$, $M_{0}$ and $L$ refer to the imaginary time, distance to the critical point, initial value of the order parameter and system size, respectively.
$x_{0}$ is the scaling dimension of $M_{0}$, satisfying $x_{0}=\theta z+\beta/\nu$ as in the classical case \cite{janssen89,diehl93,yin14}, and $\phi$ is related to the critical exponent of the quantity studied.
We consider two initial conditions $M_{0}=0$ and $M_{0}=M_{\text{sat}}$ ($M_{\text{sat}}$ being the saturate value of the order parameter) since they are both fixed points of the scaling transformation.

With $\phi=-\beta/\nu$, the $k$th moment of the order parameter follows \cite{janssen89,zbli95,zbli96,yin14}
\begin{equation}
  \label{eq:mk}
    M^{k}(\tau,g,M_{0},L)=b^{-k\beta/\nu }M^{k}(b^{-z}\tau,b^{1/\nu}g,b^{x_{0}}M_{0},b^{-1}L).
\end{equation}
At the critical point $g=0$, setting the scaling factor $b=\tau^{1/z}$, one reaches
\begin{equation}
  \label{eq:mtau}
  M^{k}(\tau,M_{0},L)=\tau^{-k\beta/\nu z}f_{M}(\tau^{x_{0}/z}M_{0},\tau^{-1/z}L),
\end{equation}
where $f_{M}$ is a scaling function related to $M^{k}$ (similar notations apply in the context).
We focus on $k=1$ (corresponding to the order parameter) and $k=2$.
In infinite systems, when $M_{0}$ is small but finite, the critical initial slip $M(\tau)\propto M_{0}\tau^{\theta}$ appears by expanding $\tau^{x_{0}/z}M_{0}$ in the short time regime $\tau \ll M_{0}^{-z/x_{0}}$ \cite{yin14}.
The presence of finite system size in Eq.~(\ref{eq:mtau}) implies that the relaxation involves another characteristic time scale $L^{z}$, which is known as the finite-size relaxation time, that controls the scale of the short time regime \cite{diehl93,zbli94}.

When $M_{0}=0$ and $g=0$, $M^{2}(\tau,L)$ obeys
\begin{equation}
  \label{eq:mL}
  M^{2}(\tau,L)=L^{-2\beta/\nu}f_{M}(\tau/L^{z}),
\end{equation}
if one choose $b=L$.
Only the finite-size relaxation time $L^{z}$ is involved in the scaling form of $M^{2}(\tau,L)$ since the initial-value time $M_{0}^{-z/x_{0}}$ diverges as $M_{0}\rightarrow 0$.

\revise{In analogy to the classical case \cite{huse89,tome98}, we consider}
\begin{equation}
  \label{eq:cm0}
  C(\tau)\equiv\lim_{M_{0}\rightarrow 0}\frac{M(\tau)}{M_{0}},
\end{equation}
\revise{which is called the imaginary time correlation function since it measures the correlation of the order parameter between the initial state and the state at $\tau$ as indicated in Appendix~\ref{app:ctau}.}
\revise{
Combining Eqs.~(\ref{eq:mk}) and (\ref{eq:cm0}), the scaling form is given by
  }
\begin{equation}
  \label{eq:cL}
  C(\tau,L)=L^{\theta z}f_{C}(\tau/L^{z}),
\end{equation}
with the scaling factor $b=L$.
In infinite systems, as indicated in Eq.~(\ref{eq:cm0}), the behavior of $C(\tau)$ is characterized by the critical initial slip exponent as well, $C(\tau)\propto\tau^{\theta}$ \cite{tome98}, while the order parameter is incapable of showing the critical initial slip when $M_{0}=0$, since symmetry restricts $M(\tau)=0$ \cite{janssen89,diehl93,zbli94,yin14}. In the presence of finite system size, $C(\tau,L)$ shows power-law increase with an exponent $\theta z$ for fixed $\tau/L^{z}$.

When $M_{0}=M_{\text{sat}}$ and $g$ varies, let us consider a quantity $Q$ related to the average sign of the order parameter, defined as \cite{oliveira92,soares97}
\begin{equation}
  Q(\tau)\equiv\braket{\text{sgn}(M(\tau))}
\end{equation}
in which $\text{sgn}$ is the sign function and $\braket{...}$ denotes non-equilibrium average. $Q(\tau)$ records the system's memory of the initial condition during the relaxation, which decays from the initial value $Q_{0}=1$ to the equilibrium value.
As argued in Refs.~\cite{oliveira92,soares97}, in Ising systems, $\phi$ is $0$, leading to the scaling form
\begin{equation}
  \label{eq:qL}
  Q(\tau,g,L)=f_{Q}(\tau/L^{z},L^{1/\nu}g),
\end{equation}
with $b=L$. It is clear that for fixed $\tau/L^{z}$, different system sizes undergo curve crossing of $Q(\tau,g,L)$ at $g=0$, in analogy to the Binder cumulants \cite{binder81_zpb,zbli95,zbli96,landau09}.

The scaling behaviors of $M^{2}(\tau,L)$, $C(\tau,L)$ and $Q(\tau,g,L)$ are useful in detecting the quantum critical point and critical exponents.

{\it(i)} With $M_{0}=M_{\text{sat}}$, the scaling form of $Q$ indicates that for fixed $\tau/L^{z}$, a series of size-dependent critical points can be extracted by performing curve crossing analysis for different system sizes \cite{landau09}. The quantum critical point is therefore obtained by extrapolating to $L\rightarrow\infty$ \cite{binder81_zpb,landau09}.

{\it (ii)} With $M_{0}=0$, when $\tau/L^{z}$ is fixed, $M^{2}(L)$ and $C(L)$ have power-law behaviors, giving the critical exponents $-2\beta/\nu$ and $\theta z$.
The requirements of small $M_{0}$ and extrapolation to $M_{0} \rightarrow 0$ to compute $\theta$ in previous investigations \cite{grassberger95,okano97,bzheng98,hpying01,albano11} is therefore avoided. Note that Eq.~(\ref{eq:mL}) also holds when $M_{0}=M_{\text{sat}}$ and $g=0$, meaning that $\beta/\nu$ can be measured in the long time regime ($\tau_{\text{cr}}\sim 1$) under this initial condition as well \cite{yin14}.

{\it (iii)} In order to get reliable estimates of the critical exponents, the leading scaling correction should be taken into account. We use the following ansatz \cite{pelissetto02,luo01,albano11}
\begin{equation}
  \label{eq:correction}
  A(L)=aL^{\sigma}(1+bL^{-\omega})
\end{equation}
to extract the critical exponents and the leading correction exponent $\omega$. In Eq.~(\ref{eq:correction}), $A(L)$ corresponds to $M^{2}(L)$ or $C(L)$ for fixed $\tau/L^{z}$ and $\sigma$ equals to $-2\beta/\nu$ and $\theta z$ accordingly. \revise{The choice of $\tau/L^{z}$ should not affect the asymptotic behavior of $A(L)$ but only bring different finite size corrections.}

The quantum STCD provides access to the critical properties while the system is relaxing towards its ground state, and it overcomes the critical slowing down problem \thirdrevise{(in the sense of physical time)}~\cite{zbli95,zbli96,yin14}, \revise{requiring much less computational efforts comparing with the traditional finite size scaling studies. }

\section{Quantum Monte Carlo Method}
\label{sec:method}
The method we employ is the projector QMC algorithm based on the stochastic series expansion (SSE), which allows us to obtain the relaxation properties of a system after evolving an imaginary time $\tau$ \cite{cdgrandi11,farhi12,cwliu13,cdgrandi13}. The idea of the QMC algorithm is to substitute the imaginary time evolution operator $e^{-\tau H}$ with its Taylor expansion series and perform importance samplings in the configuration space, which contains information needed to compute the expectation values of physical quantities.

Following Eqs.~(\ref{eq:Seq}) and (\ref{eq:formal}), the partition function $Z$ is defined as
\begin{equation}
  \label{eq:Z}
  Z=\braket{\Psi(\tau)|\Psi(\tau)} = \braket{\psi(\tau_{0})|e^{-\tau H}e^{-\tau H}|\psi(\tau_{0})},
\end{equation}
in which $\ket{\Psi(\tau)}=e^{-\tau H}\ket{\psi(\tau_{0})}$. Divide the Hamiltonian into a sum of bond operators
\begin{equation}
  \label{eq:Hb}
  H=-\sum_{b=1}^{N_{\text{op}}}{H_{b}},
\end{equation}
and series expand the exponential factor $e^{-\tau H}$ with $\beta=2\tau$, Eq.~(\ref{eq:Z}) becomes \cite{farhi12,sandvik10}
\begin{equation}
  \label{eq:ZSn}
  Z=\sum_{n}^{\infty}\sum_{S_{n}}{\braket{\psi(\tau_{0})|\frac{\beta^{n}}{n!}S_{n}|\psi(\tau_{0})}},
\end{equation}
where $S_{n}$ denotes the operator sequence of bond operators.

In order to calculate precise expectation values at imaginary time $\tau$, a binomial factor $w(n,k)=\binom{n}{k}/2^{n}$ should be inserted to Eq.~(\ref{eq:ZSn}) \cite{farhi12,sandvik10}
\begin{equation}
  \label{eq:Zwnk}
  Z=\sum_{n}^{\infty}\sum_{S_{n}}\frac{\beta^{n}}{n!}\braket{\psi(\tau_{0})|\sum_{n}^{k}{w(n,k)S_{n}^{k}}|\psi(\tau_{0})},
\end{equation}
in which $S_{n}^{k}$ is an operator sequence of length $n$ with an imaginary ``cut'' at position $k$ running through it and $w(n,k)$ satisfies
\begin{equation}
\label{sumwght}
\sum_{n_{1}=0}^{n}{w(n,n_{1})}=1, \forall n.
\end{equation}
In fact, $w(n,k)$ comes out immediately if one expand the two exponential factors independently, concatenate and sample the operator sequences together. $w(n,k)$ describes contributions of different propagations along the imaginary time direction and plays an important role when it aims at computing precise results of the states after small length of evolutions \cite{farhi12}.
As the operator sequence grows longer, the influence of $w(n,k)$ becomes smaller and the partition function reduces to Eq.~(\ref{eq:ZSn}), suggesting that measurements can be taken in the middle of the operator sequence.

The expansion order $n$ can be truncated to some maximum order $n_{\text{max}}$ without causing detectable errors \cite{sandvik10}. The expectation value of an observable $O$ is given by $\braket{O(\tau)}=\braket{\Psi(\tau)|O|\Psi(\tau)}/Z$ as in the standard SSE algorithm. In the presence of $w(n,k)$, measurements should be taken at different positions of the operator sequence, bringing different contributions to $\braket{O(\tau)}$ \cite{farhi12}.

\section{Model}
\label{sec:model}
The models we study here are the transverse field Ising models, defined by the Hamiltonian
\begin{equation}
H=-J\sum_{\braket{i,j}}^{L^d}{\sigma_{i}^{z}\sigma_{j}^{z}}-h\sum_{i}^{L^d}{\sigma_{i}^{x}},
\end{equation}
in which $\braket{i,j}$ are nearest-neighbor sites and $\sigma^{x}$ and $\sigma^{z}$ are the Pauli matrice. The couplings $J$ are chosen to be ferromagnetic interactions ($J>0$) and $h$ is a external field in the transverse direction. The order parameter is given by the magnetization $M=\braket{\sigma^{z}}$ and periodic boundary condition is used.

The 1d TFIM can be solved rigorously by mapping into the 2d classical Ising model \cite{suzuki76}. The critical point of the ordered/disordered quantum phase transition is $(h/J)_{c}=1$. The static exponents of interest here are $\beta=1/8$ and $\nu=1$ and the dynamic exponent is $z=1$ \cite{sachdev99,sondhi97,vojta03}.
Application of the STCD to the 2d classical Ising model has attracted many numerical attention \cite{grassberger95,okano97,bzheng98,hpying01,albano11}. The critical initial slip arises in the short time regime with an exponent $\theta=0.191(1)$ \cite{okano97,albano11}.
For the 1d TFIM, Ref.~\cite{yin14} extracts the critical initial slip exponent from the imaginary time relaxation, giving $\theta=0.373$, which is almost twice as large as its classical counterpart.

Owing to the fact that the 2d TFIM has no rigorous solution, many numerical efforts have been devoted to study its properties, including locating the critical point with high precision \cite{hamer00,cwliu13,blote2002,jongh1998} and extracting critical exponents \cite{pfeuty1971,jongh1998,albuquerque2010}.
For the 2d square-lattice TFIM, an exact diagonalization study on a $6\times 6$ lattice \cite{hamer00} gives the critical point $(h/J)_{c}=3.04497(18)$ and a recent work using quasi-adiabatic QMC claims a higher precision estimate $(h/J)_{c}=3.04458(7)$ \cite{cwliu13}. The dynamic exponent has an exact result $z=1$ \cite{sachdev99,sondhi97,vojta03}, while the static exponents are shared with the 3d classical Ising model since they belong to the same universality class.
A recent study of the 3d Ising universality class estimates $\beta/\nu=0.51814(5)$, \thirdrevise{which is not directly reported in Ref.~\cite{hasenbusch10} but can be obtained using the scaling relation $\beta/\nu = (1+\eta)/2$ with $\eta = 0.03627(10)$~\cite{hasenbusch10}.}
In Ref.~\cite{jaster99}, an STCD study of the 3d classical Ising model gives $\beta/\nu=0.517(2)$ along with the critical initial slip exponent $\theta=0.108(2)$.
To our knowledge, the critical initial slip of the 2d TFIM has not been studied yet.

\section{Numerical results}
\label{sec:results}
In this section, we present results of the 1d and 2d TFIMs.
Two different initial conditions $M_{0}$ both with vanishing correlation length are employed here.
We shall first use $M_{0}=M_{\text{sat}}=1$, which is a state with all spins in the up direction $\ket{\Uparrow}$.
Quenches from this ordered state to the vicinity of the critical point are imposed so that the critical coupling $h_{c}$ (for simplicity, $J=1$) can be detected through the finite size scaling analysis of $Q(\tau,g,L)$.
The other initial condition is a state with all spins aligned along the transverse direction $\ket{\Rightarrow}$ with $M_{0}=0$.
The critical initial slip can be observed during the relaxation and the exponents $\theta$, $\beta/\nu$ are extracted from the scaling behaviors of $C(\tau,L)$ and $M^{2}(\tau,L)$.
The exact value of $z=1$ \cite{sachdev99,sondhi97,vojta03} for both models is used in the scaling relations.

\revise{
In the QMC simulations, both local and cluster updates are carried out to perform efficient samplings \cite{sandvik03}. The computational time scales as $L^{d}\tau$~\cite{sandvik03} and therefore proportional to $L^{d+z}$ (as the ratio $\tau/L^{z}$ is fixed), but still, large amount of efforts required to achieve the ground state, corresponding to much larger $\tau/L^{z}$, is saved by the quantum STCD.
The convergence and autocorrelations of the QMC algorithm are discussed in Appendix~\ref{app:auto}.
}
\subsection{1d transverse field Ising model}
The 1d TFIM provides a rigorous testing ground for the QMC algorithm and the scaling for the quantum STCD in finite size systems.

\begin{figure}
  \centerline{\epsfig{file=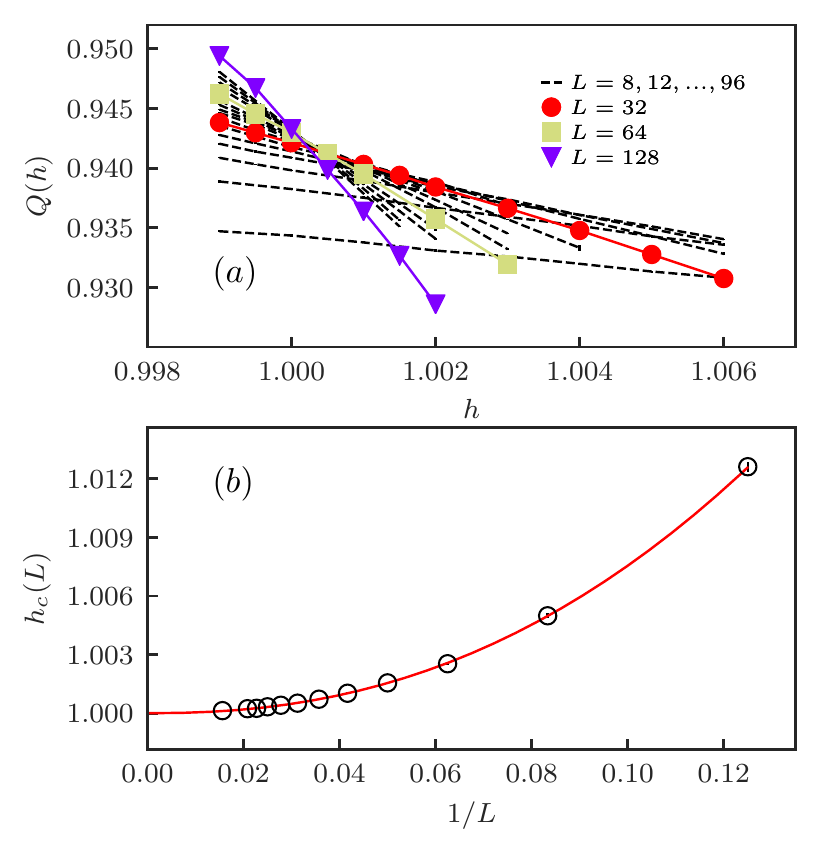,width=\columnwidth}}
  \caption{\label{fig:q1d} (Color online) Under the initial condition of $M_{0}=1$ and vanishing correlation length, curve crossing analysis for the 1d TFIM. System size varies from $L=8$ to $128$. Panel (a) shows the dependence on $h$ of $Q$ for different system sizes. The exact result $z=1$ is used and the ratio of $\tau/L$ is $1/4$. Size-dependent crossing points are determined by $Q(h)$ of $L$ and $2L$ as shown in panel (b). The solid line is a fit to the form of $h_{c}(L)=h_{c}+aL^{-\omega}$ \cite{binder81_prl} with $h_{c}=1.00003(8)$ and $\omega=2.33(1)$. To guide eyes, we use different symbols to give examples of extracting crossing points of $L$ and $2L$ and others system sizes computed are indicated by dash lines in panel (a). All data are presented with errorbars but too small to distinguish. The crossing point of $L=8$ and $L=16$ falls beyond the plot.}
\end{figure}

When the system is quenched from $\ket{\Uparrow}$ to the vicinity of the critical point, the scaling behavior of $Q(\tau,g,L)$ is described by Eq.~(\ref{eq:qL}).
We investigate $Q$'s dependence on the tunable parameter $h$ for different system sizes, as shown in the upper panel of Fig.~\ref{fig:q1d}. The argument $g$ in Eq.~(\ref{eq:qL}) is replaced by $h-h_{c}$ in order to carry out curve crossing analysis.
We fixed the evolution time at $\tau=L/4$ for $L$ from $8$ to $128$ and extract the crossing points of system size pairs of $L$ and $2L$ \cite{landau09} by fitting $Q(h)$ using polynomial forms (up to cubic terms). Panel (b) of Fig.~\ref{fig:q1d} shows the size dependence of the crossing points.
By fitting to the form of $h_{c}(L)=h_{c}+aL^{-\omega}$ \cite{binder81_prl} and extrapolating to $L\rightarrow \infty$, we obtain the critical point $h_{c}=1.00003(8)$ with $\omega=2.33(1)$, in good agreement with the exact result $h_{c}=1$ \cite{sachdev99,vojta03,sondhi97}.
\revise{
\thirdrevise{The fittings are performed by searching for parameters to have the goodness of fit $\chi^{2}$ per degree of freedom $N_{\mathrm{DOF}}$ close to $1$.}
To estimate the statistical errors of the fitting parameters, we generate different sets of data by adding Gaussian noise to the data, with the magnitude of the noise equal to the error bars of the data~\cite{sandvik10}. Repeating the fits for a large number of times, the standard deviation of the distributions give the error bars of the fitting parameters~\cite{sandvik10}. The same way of error estimating is used in the following results.
}

\begin{figure}
  \centerline{\epsfig{file=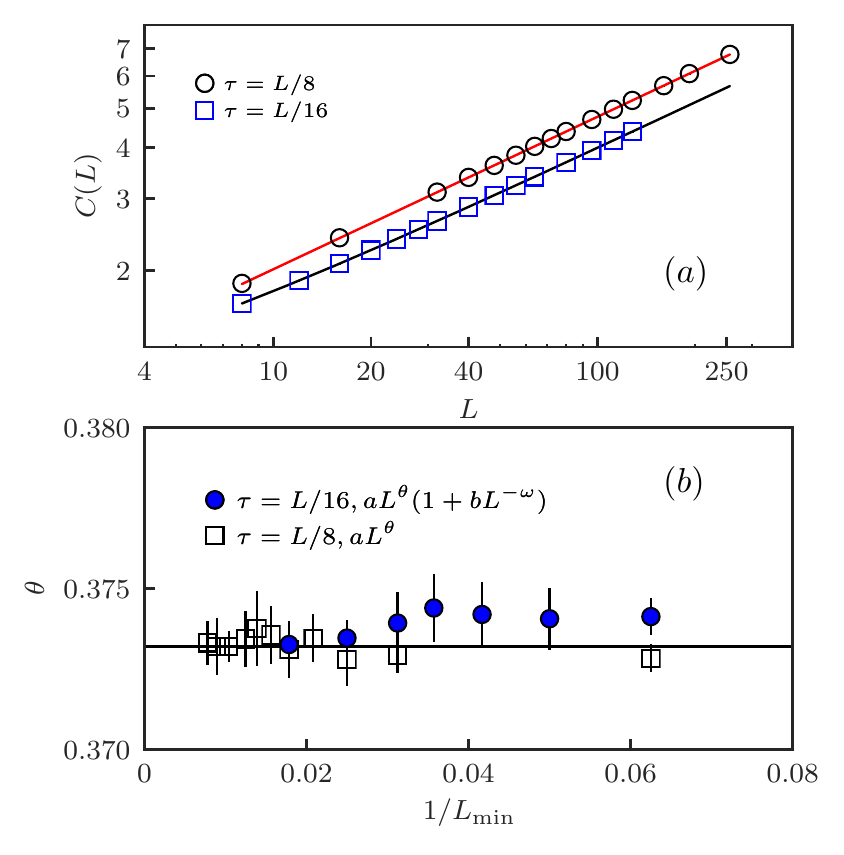,width=\columnwidth}}
  \revise{
    \caption{\label{fig:theta1d} (Color online) \thirdrevise{Panel (a): the imaginary time correlation function versus system size $L~(L=8,16,...,256)$ in a log-log plot. The initial condition is $M_{0}=0$ and two ratio of $\tau/L$ are considered.
      Solid lines are fits to power-law form in the absence/presence of the leading finite size correction for $\tau/L=1/8$ and $1/16$ respectively.
      Fitting range dependence of $\theta$ is shown in panel (b), in which the solid line represents the known value $0.373$ from Ref.~\cite{yin14}. For $\tau/L=1/8$, the critical initial slip exponent is given by $\theta=0.3734(2)$ within the fitting range from $L=32$ to $256$, along with the a prefactor $a=0.853(4)$. No finite size correction is included to the fitting form while for $\tau/L=1/16$, correction is needed to give $\theta=0.3733(6)$, along with $\omega=1.2(4)$, $a=0.71(3)$ and $b=1.0(2)$, in the fitting range $L=56\sim 128$. For a discussion see the text.
}
}
}
\end{figure}

Next we prepare the system in the state $\ket{\Rightarrow}$ and suddenly quench it to the critical point $h_{c}=1$ at $\tau=0$.
After a short microscopic period, the critical initial slip starts to emerge. We compute the imaginary time correlation function $C(\tau,L)$ according to Appendix~\ref{app:ctau} in QMC simulations.
As illustrated in Fig.~\ref{fig:theta1d}, for $\tau=L/8$, $C(L)$ behaves as a power-law form with the critical initial slip exponent $\theta$.
\thirdrevise{We perform power-law fitting using the form of $C(L)=aL^{\theta}$ to extract $\theta$. In Fig.~\ref{fig:theta1d}(b), we plot $\theta$ as a function of $1/L_{\mathrm{min}}$ ($L_{\mathrm{min}}$ being the minimal system size included into the fit) to show the dependence on the fitting range.
 Each point is obtained by performing power-law fitting with/without correction in the range from $L_{\mathrm{min}}$ to $L_{\mathrm{min}+p}$. When there is no correction included, $p=4$ is used so as to satisfy the minimum requirement of $N_{\mathrm{DOF}}=1$ while in the presence of correction, $p$ is set to $6$. As can be seen in Fig.~\ref{fig:theta1d}(b), $\theta$ only fluctuates slightly even for small system sizes. Therefore, we use the same form to fit the data within the range from $L=32$ to $256$, yielding the critical initial slip exponent $\theta=0.3734(2)$, in excellent agreement with Ref.~\cite{yin14}. No finite size correction is included since $C(L)$ is well described by the power-law form except two smallest system sizes.
  We also read off the the result $\theta=0.3733(7)$ given by the fitting range that consists of the largest system sizes $L=128$ to $256$ to provide knowledge about the magnitudes of statistical errors involved in Fig.~\ref{fig:theta1d}(b).
}
The value of $\theta$ of the 1d TFIM is in great difference from the result of the 2d classical Ising model $\theta=0.191(1)$ \cite{okano97,albano11}.

\thirdrevise{
  We also include $C(L)$ of $\tau=L/16$ in Fig.~\ref{fig:theta1d} to show the influence of different time-space ratios. Finite size correction is needed to produce good fits. As shown in Fig.~\ref{fig:theta1d}(b), the dependence on fitting range is more obvious than the case $\tau/L=1/8$. Even though $\theta$ of $L_{\mathrm{min}}=40$ and $56$ are already close, to be on the safe side, we only use the range starting from the largest $L_{\mathrm{min}}$ (i.e. $L=56\sim 128$) to represent the result of $\tau/L=1/16$, which reads $\theta=0.3733(6)$, agreeing with the result of $\tau/L=1/8$ within errorbar. The choice of $\tau/L$ does not affect the asymptotic behavior but only cause different finite size corrections. Depending on the quantity studied, optimal ratio(s) that alleviates finite size correction may exist. For $C(L)$, $\tau/L=1/8$ is close to the optimal value so that there is little finite size effect remained.
}

\begin{figure}
  \centerline{\epsfig{file=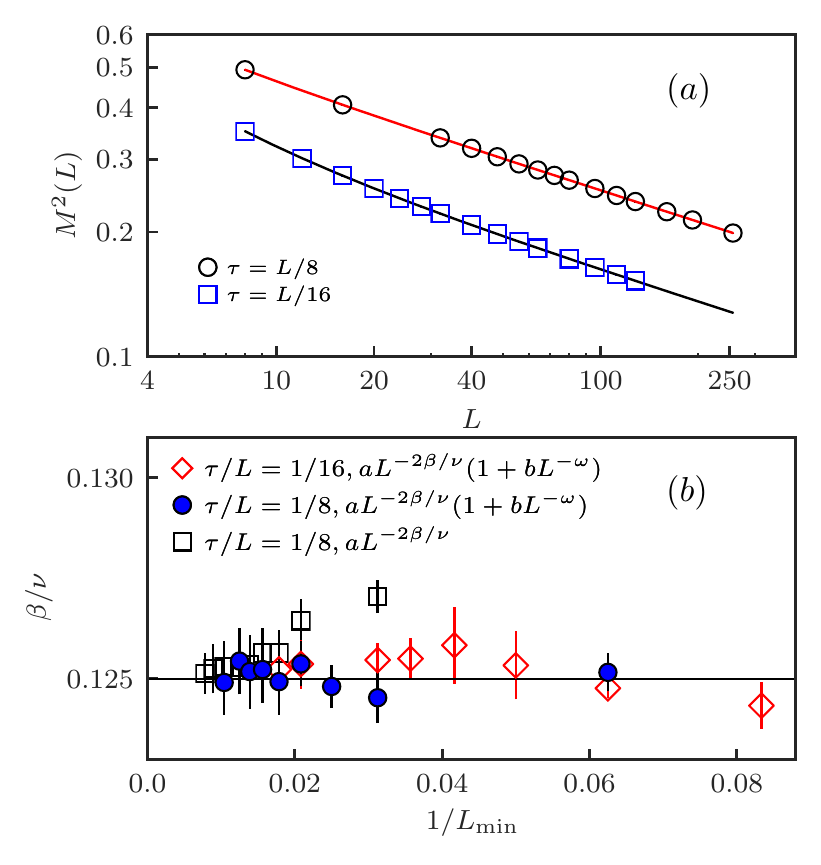,width=\columnwidth}}
  \caption{\label{fig:beta1d} (Color online) \thirdrevise{Panel (a): the second moment of the order parameter versus system size $L~(L=8,16,...,256)$ in log-log scale. The initial condition and the ratio $\tau/L$ are identical with the measurement of $C(L)$. Panel (b) compares the dependence of $\beta/\nu$ on fitting range when different fitting forms are used. Solid line marks the exact result $1/8$ \cite{sachdev99,vojta03,sondhi97}. In the presence of correction, the fits are indicated by solid lines in panel (a). For $\tau/L=1/8$, the final estimates are $\beta/\nu=0.1251(2)$, $\omega=1.07(2)$ and the prefactors being $a=0.796(2)$, $b=0.39(2)$ based on data between $L=48$ and $256$. For $\tau/L=1/16$, in the range $L=56\sim 128$, $\beta/\nu$ is given by $0.1253(5)$, along with $\omega=1.16(3)$ and $a=0.511(3)$, $b=1.9(1)$. See text for a discussion.
    }}

\end{figure}

The static critical exponent $\beta/\nu$ can be measured under the same initial condition, as shown in Fig.~\ref{fig:beta1d}. For $\tau=L/8$, \thirdrevise{we compare the result in the presence and absence of correction, both with $N_{\mathrm{DOF}}=1$, as illustrated in Fig.~\ref{fig:beta1d}(b). The exponent $\beta/\nu$ only fluctuates slightly when finite size correction is included while in the absence of correction, $\beta/\nu$ changes with varying fitting range. Using Eq.~(\ref{eq:correction}), the largest system sizes starting from $L_{\mathrm{min}}=96$ gives $\beta/\nu=0.1249(8)$. Since $\beta/\nu$ agrees within errorbar when $L_{\mathrm{min}}\ge 48$, we expand the range to $L=48 \sim 256$ and obtain $\beta/\nu=0.1251(2)$, in good agreement with the exact result $\beta/\nu=1/8$ \cite{sachdev99,vojta03,sondhi97}.
}

\thirdrevise{
  Likewise, we plot $M^{2}(L)$ for $\tau/L=1/16$ in Fig.~\ref{fig:beta1d}(a) and the dependence of $\beta/\nu$ on fitting range in panel (b). Even in the presence of correction, $\beta/\nu$ only becomes close when $L_{\mathrm{min}}\ge 48$. To be safe, we use the result of the largest $L_{\mathrm{min}}=56$ to represent the final result $\beta/\nu=0.1253(5)$, consistent with the result of $\tau/L=1/8$ and the exact value as well. Figure \ref{fig:beta1d}(b) also implies that to get more reliable estimate, larger systems should be involved.
}

\thirdrevise{We therefore arrive at the final estimates of $\theta$ and $\beta/\nu$ for the 1d TFIM
  \begin{eqnarray*}
    \label{eq:final1d}
    \theta &=& 0.3734(2), \\
    \beta/\nu &=& 0.1251(2),
  \end{eqnarray*}
  based on analysis on data of $\tau/L=1/8$ within the fitting range $L=32\sim 256$ and $L=48\sim 256$, respectively.  Note that the errors here for the exponents $\theta$, $\beta/\nu$(as well as $\omega$, $a$ and $b$) are only statistical variations among the data points in a given range.

}
Our results agree nicely with the rigorous solutions \cite{sondhi97,sachdev99,vojta03} and the numerical results \cite{yin14}, demonstrating the validity of the quantum STCD and the QMC algorithm in detecting critical properties.

\subsection{2d transverse field Ising model}
In the following, we apply the quantum STCD to the 2d TFIM.

\begin{figure}
  \centerline{\epsfig{file=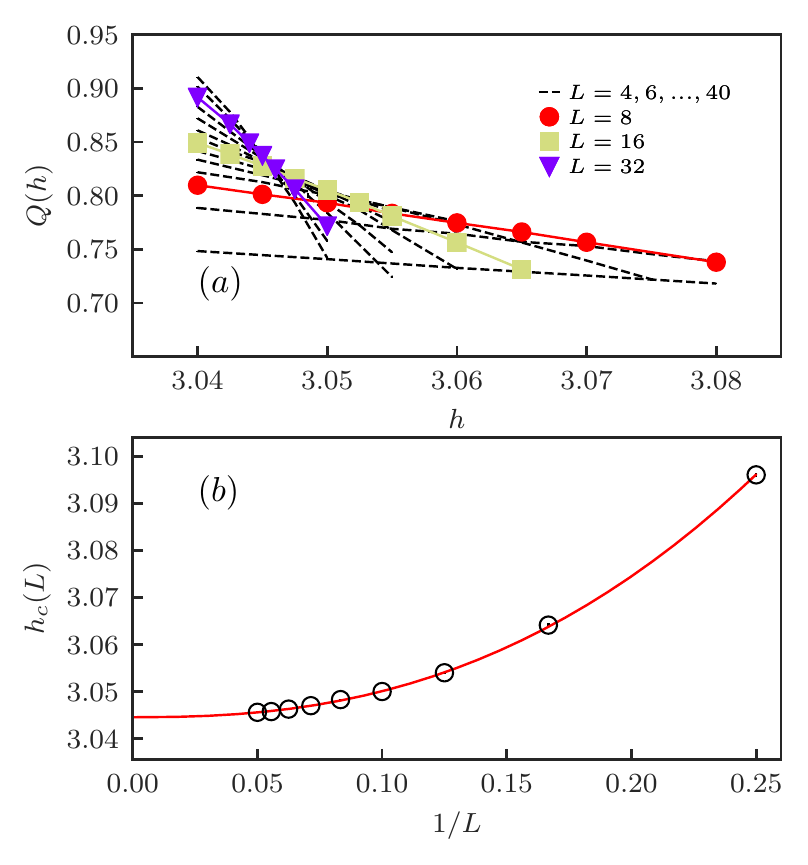,width=\columnwidth}}
  \caption{\label{fig:q2d} (Color online) Curve crossing analysis for the 2d TFIM with $M_{0}=1$. System size varies from $L=4$ to $40$. Panel (a) shows dependence on $h$ of $Q$ for different system sizes. The exact result $z=1$ is used and the ratio $\tau/L$ is $1/4$. Size-dependent crossing points are determined by $Q(h)$ of $L$ and $2L$ as shown in panel (b). Solid line is a fit to the form of $h_{c}(L)=h_{c}+aL^{-\omega}$ \cite{binder81_prl} with $h_{c}=3.04451(7)$ and $\omega=2.44(3)$. Similar to Fig.~\ref{fig:q1d}, three $Q(h)$ curves are highlighted while others are indicated by dash lines. The crossing point of $L=4$ and $8$ is too far away to show in the plot.}
\end{figure}

As in the 1d case, we shall first determine the critical point by carrying out curve crossing analysis of $Q(h,L)$ with $\tau=L/4$ and $M_{0}=1$, as illustrated in Fig.~\ref{fig:q2d}. Crossing points of different system size pairs of $L$ and $2L$ are extracted using polynomial fits (up to cubic terms) of the $Q(h,L)$ \cite{landau09}.
Using the form of $h_{c}(L)=h_{c}+aL^{-\omega}$ \cite{binder81_zpb}, we estimate the critical coupling $h_{c}=3.04451(7)$, in good agreement with the result in Ref.~\cite{cwliu13}, \thirdrevise{in which the calculations are carried out on systems of similar sizes to those in this study}.

\begin{figure}
  \centerline{\epsfig{file=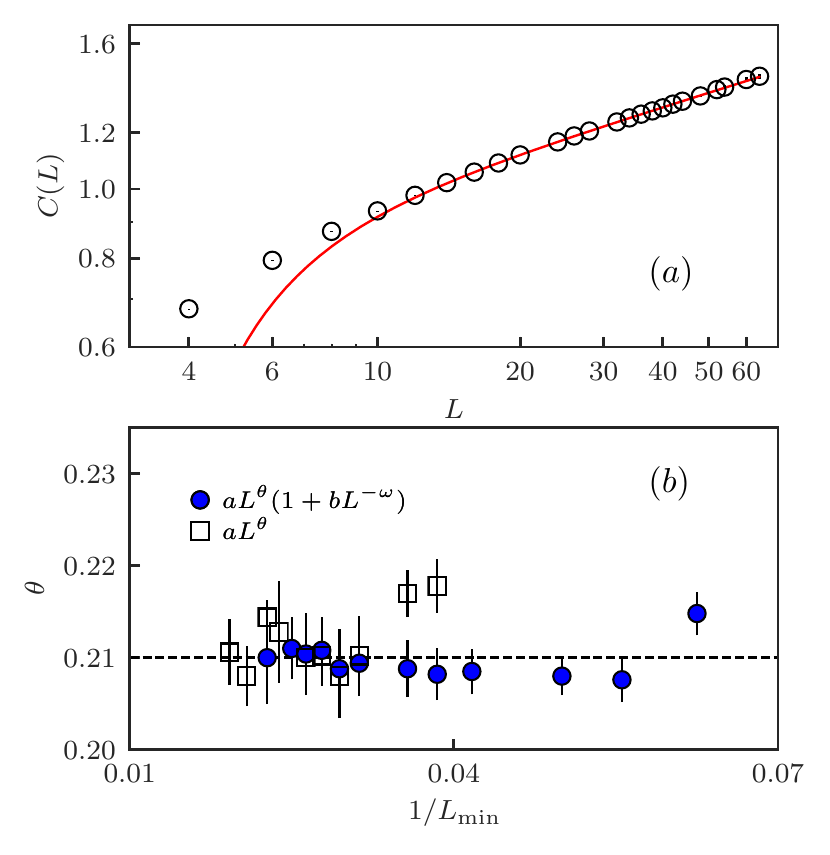,width=\columnwidth}}
  \caption{\label{fig:theta2d} (Color online) \thirdrevise{Panel (a): the imaginary time correlation function versus system size $L (L=4,6,...,64)$ in a log-log plot. The evolution time is $\tau=L/4$ and the initial condition is $M_{0}=0$. The solid line is a fit to the form of $C(L)=aL^{\theta}(1+bL^{-\omega})$ in the range from $L=24$ to $L=64$. The critical initial slip exponent is determined as $\theta=0.209(4)$ with the correction exponent $\omega=2.25(6)$ and $a=0.601(5)$, $b=-9.4(6)$.
    Panel(b): dependence of $\theta$ on the fitting range estimated in different fitting ranges using a power-law form with/without correction. The dashed line marks the result of $L_{\mathrm{min}}=44$. See text for a discussion.}}
\end{figure}

When the system is suddenly quenched from the state $\ket{\Rightarrow}$ to the critical point $h_{c}=3.04451$, as expected, the initial increase of $C(\tau,L)$ arises after a transient microscopic period.
\thirdrevise{In Fig.~\ref{fig:theta2d}, we plot $C(L)$ for $\tau=L/4$ and compare the exponent $\theta$ obtained using different fitting forms.
  When there is no correction, the power-law form only captures the behavior of $C(L)$ for large systems and small system sizes should be discarded in order to get good fits. With the leading correction, the results are improved and become converged (within errorbar) for large $L_{\mathrm{min}}$. For comparison, we marks the result $\theta=0.210(7)$ of $L_{\mathrm{min}}=44$ in the presence of correction using a dashed line in Fig.~\ref{fig:theta2d}(b). Since $\theta$ converges when $L_{\mathrm{min}}\ge 24$, we use Eq.~(\ref{eq:correction}) to fit data in the range $L=24\sim 64$, yielding $\theta=0.209(4)$, along with the leading correction exponent $\omega=2.25(6)$.}
\thirdrevise{The value of $\theta$ is again distinct from its classical counterpart, the 3d classical Ising model, $\theta=0.108(2)$ \cite{jaster99}.


}
In both 1d and 2d, the quantum models have a critical initial slip exponent approximately twice as large as their classical counterparts. In fact, there is no reason to expect the critical initial slip exponent are in the $d$ to $d+z$ classical/quantum correspondence. Though both dissipative, the dynamics governing the two cases are essentially different. The quantum systems are described by the Schr\"{o}dinger's equation while the classical systems follow the master equation, with different dynamic exponents $z$ \cite{yin14,zhang14}. Specifically, the dynamic exponent of the 1d and 2d TFIMs both equal to $1$ \cite{sachdev99,sondhi97,vojta03} while in the 2d and 3d classical Ising models, \revise{the Metropolis dynamics gives} $z$ equals to $2.1667(5)$ \cite{nightingale00} and $2.042(6)$ \cite{jaster99}, respectively. The critical initial slip exponent $\theta$ is dependent on the dynamical equation \cite{yin14,zhang14} and different dynamic exponents $z$ can lead to different values of $\theta$.
However, whether or not the similar ratios of the $\theta$'s between the quantum models and their classical counterparts are coincidental may need further investigations.

\begin{figure}
  \centerline{\epsfig{file=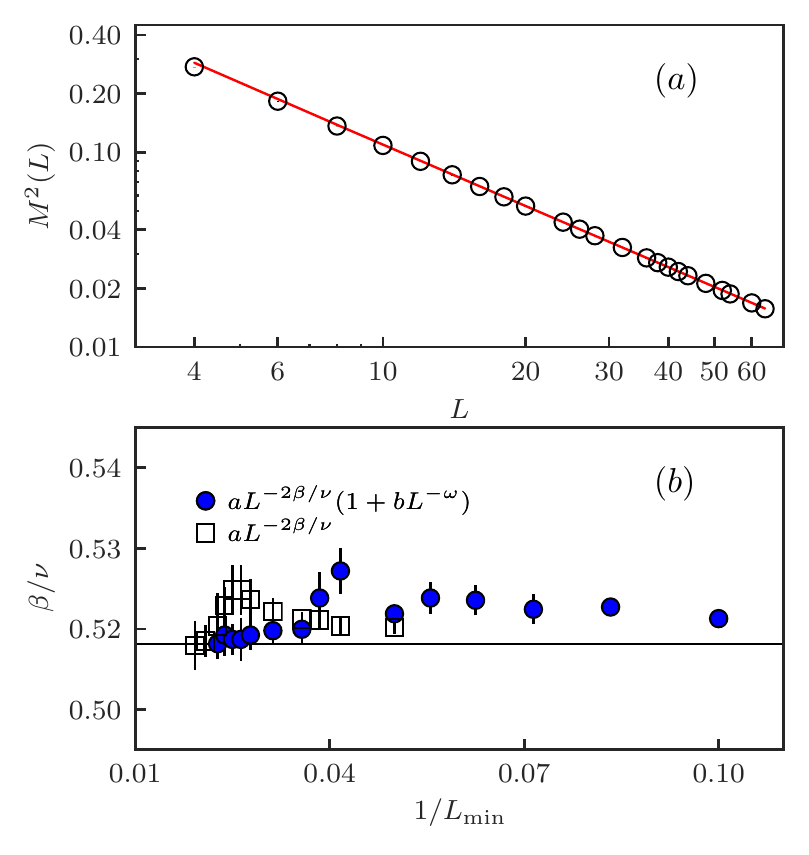,width=\columnwidth}}
  \caption{\label{fig:beta2d} (Color online) \thirdrevise{Panel (a): the second moment of the order parameter versus system size $L$ in log-log scale. The initial condition is $M_{0}=0$ and measurements are taken at $\tau=L/4$. The solid line is a fit to the form of $M^{2}(L)=aL^{-2\beta/\nu}(1+bL^{-\omega})$ with the fitting range from $L=32$ to $L=64$. The exponent is determined to be $\beta/\nu=0.518(1)$ with the leading correction exponent $\omega=0.83(2)$ and the prefactors being $a=1.17(1)$, $b=0.13(2)$.
Panel (b): $\beta/\nu$ obtained by different fitting forms in varying range. The solid line indicates the value from Ref.~\cite{hasenbusch10}. See text for a discussion.
    }
  }
\end{figure}

Next, we measure the static critical exponent $\beta/\nu$ in the same way, as seen in Fig.~\ref{fig:beta2d}. The initial condition and evolution time are the same with the measurement of $C(L)$.
\thirdrevise{Panle (b) of Fig.~\ref{fig:beta2d} compares the exponent $\beta/\nu$ obtained using different fitting forms. In the absence of correction, $\beta/\nu$ strongly depends on the fitting range. Including the leading correction produces steady results when $L_{\mathrm{min}}\ge 32$, among which, we read off the exponent $\beta/\nu=0.518(2)$ given by the largest $L_{\mathrm{min}}=44$. Expanding the fitting range to $L=32\sim 64$, we obtained $\beta/\nu=0.518(1)$, agreeing well with with the 3d classical Ising model result $\beta/\nu=0.51814(5)$ \cite{hasenbusch10}.
}

\thirdrevise{We reach our final estimates of $\theta$ and $\beta/\nu$ for the 2d TFIM
  \begin{eqnarray*}
    \label{eq:final}
    \theta &=& 0.209(4), \\
    \beta/\nu &=& 0.518(1),
  \end{eqnarray*}
  using the results in the range $L=24\sim 64$ and $L=32\sim 64$, correspondingly. Again, the errors only represent statistical variations in the fitting range.
}
\fourthrevise{As can be seen in Figs.~\ref{fig:theta2d} and \ref{fig:beta2d}, with the leading finite size correction, small system sizes down to $L=24$ and $32$ (for $C(L)$ and $M^{2}(L)$, respectively) can be included in the fitting function but subleading finite size corrections are still needed in order to take all sizes into consideration. }

\section{Summary}
\label{sec:summary}
In summary, we studied the quantum STCD in imaginary relaxation of finite size systems. By imposing different quench protocols on equilibrium systems, critical properties can be detected from the universal scaling behaviors and the critical slowing down is avoided \cite{diehl93,zbli94,zbli95,zbli96,yin14}. The method is applied to the 1d and 2d TFIMs using QMC simulations.

We located the quantum critical points of the 1d and 2d TFIMs with high precision, which are $h_{c}=1.00003(8)$ and $3.04451(7)$, respectively. The result of the 1d TFIM agrees excellently with the exact result \cite{sachdev99,sondhi97,vojta03} and the 2d case is in good agreement with the critical coupling ratio given in Ref.~\cite{cwliu13}.
The emergence of the critical initial slip are observed in both models.
In 1d, finite size scaling analysis \thirdrevise{on data between $L=32\sim 256$} gives the critical initial slip exponent $\theta=0.3734(2)$, in excellent agreement with Ref.~\cite{yin14}, but approximately twice as large as its classical counterpart $\theta=0.191(1)$ \cite{okano97,albano11}.
In 2d, though predicted from the quantum STCD, no previous numerical observation is reported.
Remarkably, our calculations are able to capture the critical initial slip exponent of the 2d TFIM $\theta=0.209(4)$ \thirdrevise{within the range $L=24\sim 64$} , which is again different from the value $0.108(2)$ of the 3d classical Ising model \cite{jaster99}.
Different critical initial exponents are actually expected in the dissipative relaxation processes of quantum and classical systems. Even though one can find equations that bear similar evolution properties in both cases, the dynamics are essentially different. In classical systems, the master equation governs the evolution after a sudden quench while in quantum systems, the Schr\"{o}dinger equation is responsible for the imaginary time evolution \cite{yin14,zhang14}.
In addition, we obtained the static exponent $\beta/\nu=0.1251(2)$ of the 1d TFIM \thirdrevise{using data within $L=48$ and $256$}, agree well with the exact result $1/8$ \cite{sachdev99,sondhi97,vojta03}. For the 2d TFIM, we estimate $\beta/\nu$ to be $0.518(1)$ \thirdrevise{in the range from $L=32$ to $64$}, in good agreement with a recent high-precision numerical estimate $0.51814(5)$ \cite{hasenbusch10}. \fourthrevise{It is also helpful to notice that, in the 2d case, the non-monotonic behavior of $\beta/\nu$ as a function of $L_{\mathrm{min}}$ indicates that there are potential corrections to the scaling forms introduced by short time effect in addition to finite size effect. In general, when $\tau/L^{z}\ll 1$, finite size effect plays a minor role while for $\tau/L^{z}\gg 1$, the scaling forms are reduced to the static case. In our calculations, the time-space ratio $\tau/L^{z}$ is fixed at values that both effects can come into play while we have only considered finite size corrections at present. Even though the short time effect can be taken into account by including leading, subleading finite size correction terms to the scaling forms (since $\tau/L^{z}$ is fixed) and we have shown that consistent asymptotic behaviors can already be obtained with only the leading finite size correction included, different ratios should indeed be considered in order to analyze the corrections caused by finite size and short time effect to the scaling forms in further studies.}
Our results manifest the capability of the quantum STCD to determine critical properties and indicate broader applications of the method to many other ``sign-problem'' free models.

\section*{acknowledgement}
We would like to thank A. Sandvik, S. Capponi and J. Marino for their valuable discussions and critical reading this manuscript. \thirdrevise{We also thank an anonymous referee for his/her sugguestions on data analysis.}
\revise{YRS thanks P. Weinberg, L. Wang and N. Xu for their useful discussions.}
This project is supported by NKRDPC-2017YFA0206203, NSFC-11574404, NSFC-11275279, NSFG-2015A030313176, Special Program for Applied Research on Super Computation of the NSFC-Guangdong Joint Fund (the second phase), National Supercomputer Center in Guangzhou, and Leading Talent Program of Guangdong Special Projects.


\appendix
\renewcommand\thefigure{\thesection.\arabic{figure}}
\setcounter{figure}{0}
\section{Measurement of the imaginary time correlation function}
\label{app:ctau}
Here we discuss the measurement of the imaginary time correlation function $C(\tau)$ in QMC simulations.

In the standard $\sigma^{z}$ basis, let us consider a state with a given magnetization $M_{0}$ at $\tau_{0}$
\begin{equation}
  \label{eq:psim0}
  \ket{\psi(\tau_{0})}=\sum_{z}{\sqrt{f(M_{0})}\ket{z}},
\end{equation}
which a superposition of all $2^{N}$ ($N=L^{d}$ being the total particle number) basis states with different amplitudes. The coefficients related to $M_{0}$ are given by
\begin{equation}
f(M_{0})=\prod_{j}{\frac{1}{2}(1+M_{0}\sigma^{z}_{j})},
\end{equation}
if one consider constructing a state site by site \cite{tome98}. \revise{For small $M_{0}$, up} to linear terms in $M_{0}$
\begin{equation}
f(M_{0})=\frac{1}{2^{N}}(1+M_{0}\sum_{j}{\sigma^{z}_{j}}).
\label{pm0}
\end{equation}
The partition function in Eq.~(\ref{eq:Z}) becomes
\begin{equation}
Z=\frac{1}{2^{N}}\sum_{z_{1},z_{2}}{\braket{z_{1}|e^{-\tau H}\sqrt{f(M_{0})f(M_{0})}e^{-\tau H}|z_{2}}},
\end{equation}
Plugging in Eq.~(\ref{pm0}), one arrives at
\begin{equation}
Z=\frac{1}{2^{N}}\sum_{z_{1},z_{2}}{\braket{z_{1}|e^{-\tau H}(1+M_{0}\overline{M})e^{-\tau H}|z_{2}}},
\end{equation}
where $\overline{M}$ equals to $[\sum_{i}{\sigma_{i}^{z}}+\sum_{j}{\sigma_{j}^{z}}]/2$. The indices $i$ and $j$ correspond to sum over the basis state $\ket{z_{1}}$ and $\ket{z_{2}}$, respectively. Therefore, the magnetization at time $\tau$ is given by
\begin{eqnarray}
M(\tau) &=& \frac{1}{Z}\sum_{z_{1}z_{2}}{\frac{1}{2^{N}}\braket{z_{1}|e^{-\tau H}Me^{-\tau H}|z_{2}}} \nonumber\\
   &+&\frac{1}{Z}\sum_{z_{1}z_{2}}{\frac{M_{0}}{2^{N}}\braket{ z_{1}|e^{-\tau H}M\overline{M}e^{-\tau H}|z_{2}}},
\end{eqnarray}
in which $M=\sum_{i}{\sigma^{z}_{i}(\tau)}/N$. The first term on r.h.s vanishes due to up-down symmetry. Divide $M(\tau)$ by $M_{0}$ before taking the limit $M_{0}\rightarrow 0$, one gets
\begin{eqnarray}\label{limm0}
   \lim_{M_{0}\rightarrow 0}\frac{M(\tau)}{M_{0}}&=&
\frac{1}{Z}\sum_{z_{1}z_{2}}{\frac{1}{2^{N}}
\braket{z_{1}|e^{-\tau H}\overline{M}Me^{-\tau H}|z_{2}}}\nonumber\\
&=&\frac{1}{N}\braket{\overline{M}[\sum_{i}{\sigma_{i}^{z}(\tau)}]},
\end{eqnarray}
where $\overline{M}$ is averaged over the two initial states $\ket{z_{1}}$ and $\ket{z_{2}}$ on the boundaries of sampling space in QMC simulations.

\secondrevise{
As indicated by the second line of Eq.~(\ref{limm0}), $C(\tau)$ measures the correlation between the initial state and the state at imaginary time $\tau$.
$C(\tau)$ actually corresponds to $\braket{\sigma_{k}^{z}(0)\sigma_{-k}^{z}(\tau)}$ with momentum $k=0$~\cite{huse89}, where $\sigma_{k}^{z}(\tau)$ is the Fourier transform of the spins at $\tau$ and $\sigma_{k}^{z}(0)$ is averaged over $\ket{z_{1}}$ and $\ket{z_{2}}$ at $\tau=0$. For small $k$, the scaling form of $C(\tau)$ can be generalized by including the scale transformation of the momentum $k$, which reads $k^{\prime}=bk$. However, for large $k$, the fluctuation modes are far away from the low energy regime, hence, the behavior is not described by the scaling form. Therefore, we restrict our calculations to the $k=0$ sector and use a disordered initial state with $M_{0}=0$ and vanishing correlation length, as required by the quantum STCD for the scaling forms to apply. The critical initial slip exponent can be conveniently extracted without suffering from the requirements of small but finite magnetization and extrapolation to $M_{0}\rightarrow 0$.
}

\revise{
\section{Convergence and autocorrelations}
\label{app:auto}
In this section, we present the convergence and the autocorrelations for the 1d TFIM to show the performance of the projector QMC algorithm.

\begin{figure}
  \centerline{\epsfig{file=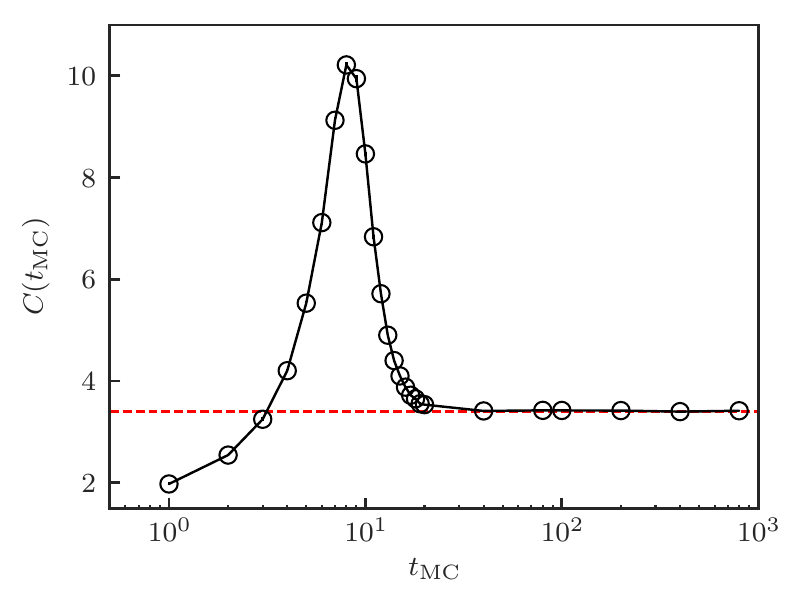,width=\columnwidth}}
  \caption{\label{fig:conv} (Color online) Convergence of $C(t_{\text{MC}})$ for $\tau=4$, $L=128$. Each point is averaged over $10^{3}$ to $10^{4}$ independent simulations. The dash line indicates the result of $t_{\text{MC}}=800$.
  }
\end{figure}

To distinguish from the real time $t$ in the main text, we use $t_{\text{MC}}$ to refer to the Monte Carlo~(MC) steps in the following. One MC step consists of a full sweep of single operator updates followed by constructions and updates of the clusters, which are flipped with probability $1/2$~\cite{sandvik03}. The details of sampling are described in Refs.~\cite{sandvik03,sandvik10}.

As a convergence test, we compute the imaginary time correlation as a function of $t_{\text{MC}}$ for $\tau=4$ and $L=128$ with an initially disordered state \thirdrevise{(which is $\ket{\Rightarrow}$)}, shown in Fig.~\ref{fig:conv}. $C(t_{\text{MC}})$ converges to the equilibrium result rapidly after $t_{\text{MC}}\ge 40$. Each data point in Fig.~\ref{fig:conv} is averaged over $10^{3}$ to $10^{4}$ independent simulations. The growth of $C(t_{\text{MC}})$ at short $t_{\text{MC}}$ corresponds to the increases of the cut-off expansion order $n_{\text{max}}$ (in Sec.~\ref{sec:method}) and the number of flipped clusters at the beginning of the simulations. As $t_{\text{MC}}$ gets longer, $C(t_{\text{MC}})$ approaches to its equilibrium value.

To accelerate equilibration, we use a $\tau$-doubling trick, which is very similar to the $\beta$-doubling in standard SSE algorithm~\cite{sandvik02}. For a given evolution time $\tau_{\text{max}}$, the $\tau$-doubling process starts from sampling a much shorter time $\tau=\tau_{\text{min}}$ with $t_{\text{MC}}$ (typically a few hundreds) and doubles the evolution time to $\tau=2\tau_{\text{min}}$ by sampling the doubled operator sequence $S_{2n}$~\cite{sandvik02}. Repeating the doubling until $\tau=\tau_{\text{max}}$, the initial configuration is almost equilirated since it has already rather long history at shorter times~\cite{sandvik02}. For the data presented in the main text, to be safe, at the last doubling, we use $t_{\text{MC}}=10^{4} \sim 10^{5}$ to perform the equilibrations and measurements for $50 \sim 200$ bins (\thirdrevise{one bin consists of $10^{4} \sim 10^{5}$ MC steps, and the bin number depends on the system sizes and the quantity studied}). The statistical errors are estimated as one standard deviation of the average of the bin averages, which are reliable only when $t_{\text{MC}}$ is much longer than the autocorrelation time so that the bin averages can be regarded as statistically independent.

\begin{figure}
  \centerline{\epsfig{file=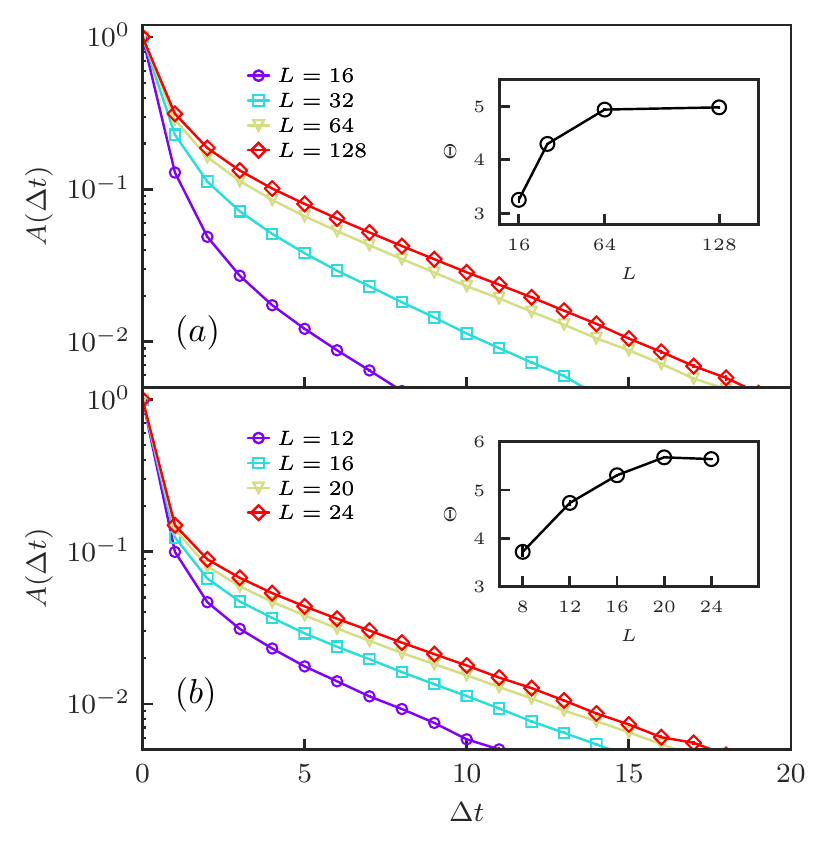,width=\columnwidth}}
  \caption{\label{fig:aucr} (Color online) Autocorrelation functions of $C(\tau,L)$ for s(a) the 1d TFIM at $\tau=16$ with $L=16,32,64,128$ and (b) the 2d TFIM at $\tau=4$ with $L=12,16,20,24$ in a log-linear scale. The size-dependent autocorrelation times are extracted by the long time exponential decay behaviors, as shown in the insets.
  }
\end{figure}

Next we discuss autocorrelation function, which, for an observable $O$, is defined as \cite{landau09}
\begin{equation}
  \label{eq:autocor}
  A(\Delta t)=\frac{\braket{O_{i+\Delta t}O_{i}}-\braket{O_{i}}^{2}}{\braket{O_{i}^{2}}-\braket{O_{i}}^{2}},
\end{equation}
where the averages are over the MC time index $i$ and $\Delta t$ is the separation of the two configurations. $A(0)$ is normalized to $1$ and when $\Delta t$ is large, $A(\Delta t)$ approaches $0$ exponentially
\begin{equation}
  \label{eq:auexp}
  A(\Delta t)\rightarrow e^{-\Delta t/\Theta},
\end{equation}
where $\Theta$ is called the autocorrelation time. Figure~\ref{fig:aucr}(a) shows the autocorrelation functions of $C(\tau,L)$ at $\tau=16$ for the 1d TFIM with system sizes $L=16,32,64$ and $128$ under the initial condition of $M_{0}=0$ and a very short correlation length. The autocorrelation times are extracted by fitting the long time behaviors of $A(\Delta t)$ using the form in Eq.~(\ref{eq:auexp}), as shown in the inset of Fig.~\ref{fig:aucr}(a). \thirdrevise{In fact, we expect similar behaviors of $A(\Delta t)$ and $\Theta$ for the 2d TFIM, as shown in Fig.~\ref{fig:aucr}(b), in which $A(\Delta t)$ is computed for $C(\tau,L)$ with $\tau=4$ and $L=12,16,20,24$.} Since the initial state has a vanishing correlation length, after a short length of imaginary time evolution (not sufficient to reach the ground state), the correlation length is still finite in the short time regime.
When the system size grows larger than the correlation length, the autocorrelation time becomes insensitive to the system size, indicating that the critical slowing down problem is overcome. The autocorrelation times are only up to a few MC steps so that in the simulations, the number of steps we used $t_{\text{MC}}=10^{4}$ to $10^{5}$ is sufficient to generate statistically independent configurations.

}
\null\vskip-8mm
\bibliography{st}
\end{document}